\documentclass[10pt, conference, compsocconf]{IEEEtran}
\usepackage{slashbox}
\usepackage{dsfont,graphics,epsfig,psfrag,stmaryrd,amsmath,amsfonts,amssymb,mathrsfs,txfonts,booktabs,times} 
\usepackage{amsmath}
\usepackage{graphicx}
\usepackage{amsfonts}
\usepackage{amssymb}
\usepackage{multirow}
\usepackage{alltt}
\usepackage{booktabs}
\usepackage{stmaryrd}
\usepackage{dsfont} 
\usepackage{color} 
\usepackage[standard]{ntheorem}
\usepackage[font=footnotesize]{subfig}
\usepackage{array,algorithm2e,algorithmic}

\usepackage[width=18.2cm,height=25.7cm]{geometry}
\usepackage{epstopdf}

\ifCLASSINFOpdf
\else
\fi

\begin{document}
\title{An optimization technique on pseudorandom generators based on chaotic iterations}

\author{\IEEEauthorblockN{Jacques M. Bahi,
Xiaole Fang, and
Christophe Guyeux*}
\IEEEauthorblockA{FEMTO-ST Institute, UMR 6174 CNRS\\
University of Franche-Comt\'{e},
Besan\c con, France\\ Email:\{jacques.bahi, xiaole.fang,
christophe.guyeux\}@univ-fcomte.fr }\\ }

%

\maketitle

\begin{abstract}
Internet communication systems involving cryptography and data hiding often require billions of random numbers. In addition to the speed of the algorithm, the quality of the pseudo-random number generator and the ease of its implementation are common practical aspects. In this work we will discuss how to improve the quality of random numbers independently from their
generation algorithm. We propose an additional implementation technique in order to take advantage of some chaotic properties. The statistical quality of our solution stems from some well-defined discrete chaotic iterations that satisfy the reputed Devaney's definition of chaos, namely
the chaotic iterations technique. Pursuing recent researches published in the previous International Conference on Evolving Internet (Internet 09, 10, and 11), three methods to build pseudorandom generators by using chaotic iterations are recalled. Using standard criteria named NIST and DieHARD (some famous batteries of tests), we will show that the proposed technique can improve the statistical properties of a large variety of defective pseudorandom generators, and that the issues raised by statistical tests decrease when the power of chaotic iterations increase.

\end{abstract}

\begin{IEEEkeywords}
Internet security; Pseudorandom Sequences; Statistical Tests; Discrete Chaotic Iterations; Topological Chaos.
\end{IEEEkeywords}

\IEEEpeerreviewmaketitle

\section{Introduction}

Chaos has recently attracted more and more interests from researchers in the fields of mathematics, physics, and computer engineering, among other things due to its connection with randomness and complexity~\cite{Behnia20113455,Guyeux10}.
In particular, various research works have recently regarded the possibility to use chaos in random number generation for Internet security.
Indeed, the security of data exchanged through the Internet is highly dependent from the quality
of the pseudorandom number generators (PRNGs) used into its protocols. These PRNGs are everywhere
in any secure Internet communication: in the keys generation of any asymmetric cryptosystem, in
the production of any keystream (symmetric cryptosystem), the generation of nonce, in the
keys for keyed hash functions, and so on.

Numerous pseudorandom number generators already exist, but they are either secure but slow, or
fast but insecure.
This is why the idea to mix secure and fast PRNGs, to take benefits from their respective qualities, has emerged these last years~\cite{Guyeux10,guyeux10ter}.
Chaotic dynamical systems appear as good candidates to achieve this mixture for optimization.
Indeed, chaotic systems have many advantages as unpredictability or disorder-like, which are required in building complex sequences~\cite{Hu20092286,DeMicco20083373}. This is why chaos has been applied to secure optical communications~\cite{Larger10}.
But chaotic systems of real-number or infinite bit representation realized in finite computing precision lead to short cycle length, non-ideal distribution, and other deflation of this kind.
This is the reason of that chaotic systems on an infinite space of integers have been looked for these last years, leading to the proposition to use chaotic iterations (CIs) techniques to reach the desired goals.
More precisely, we have proposed in INTERNET 2009~\cite{bgw09:ip} to mix two given PRNGs by using chaotic iterations, being some particular kind of discrete iterations of a vectorial Boolean function.
This first proposal has been improved in INTERNET 2010~\cite{wbg10:ip} and INTERNET 2011~\cite{bcgw11:ip}, to obtain a new family of statistically perfect and fast PRNGs.
A short overview of these previous researches is given thereafter.

In~\cite{Guyeux10}, CIs have been proven to be a suitable tool for fast computing iterative algorithms on integers satisfying the topological chaotic property, as it has been defined by Devaney~\cite{Dev89}.
A first way to mix two given generators by using these chaotic iterations, called Old CIPRNGs, has been proposed in Internet 09 \cite{bgw09:ip} and further investigated in~\cite{bgw10:ip,guyeux09,guyeuxTaiwan10}. It was chaotic and able to pass the most stringent batteries of tests, even if the inputted generators were defective.
 This claim has been verified experimentally, by evaluating the scores of the logistic map, XORshift, and ISAAC generators through these batteries, when considering them alone or after chaotic iterations.
Then, in \cite{wbg10:ip}, a new version of this family has been proposed.
This ``New CIPRNG'' family uses a decimation of strategies leading to the improvement of both speed and statistical qualities. Finally, efficient implementations on  GPU using a last family called Xor CIPRNG have been designed in~\cite{arxivRCCGPCH}, showing that a very large quantity of pseudorandom numbers can be generated per second (about 20 Gsamples/s).

In this paper, the statistical analysis of the three methods mentioned above are carried out systematically, and the results are discussed.
Indeed PRNGs are often based on modular arithmetic, logical operations like bitwise exclusive or (XOR), and on circular shifts of
bit vectors.
However the security level of some PRNGs of this kind has been revealed inadequate by today's standards.
Since different biased generators can possibly have their own side effects when inputted into our mixed generators, it is normal to enlarge the set of tested inputted PRNGs, to determine if the observed improvement still remains.
We will thus show in this research work that the intended statistical improvement is really effective for all of these most famous generators.

The remainder of this paper is organized in the following way. In Section~\ref{Basic recalls}, some basic definitions concerning chaotic iterations are recalled. Then, four major classes of general PRNGs are presented in Section~\ref{The generation of pseudo-random sequence}. Section~\ref{Security analysis} is devoted to two famous statistical tests suites. In Section~\ref{Results and discussion}, various tests are passed with a goal to achieve a statistical comparison between our CIPRNGs and other existing generators. The paper ends with a conclusion and intended future work.

\section{Chaotic Iterations Applied to PRNGs}
\label{Basic recalls}

In this section, we describe the CIPRNG implementation techniques that can improve the statistical properties of any generator. They all are based on CIs, which are defined below.

\subsection{Notations}
\begin{tabular}{@{}c@{}@{}l@{}}
$S^{n}$ & $\rightarrow$ the $n^{th}$ term of a sequence $S=(S^{1},S^{2},\hdots)$ \\
$v_{i}$ & $\rightarrow$ the $i^{th}$ component of a vector $v=(v_{1},\hdots, v_n)$\\
$f^{k}$ & $\rightarrow$ $k^{th}$ composition of a function $f$ \\
$\emph{strategy}$~ & $\rightarrow$ a sequence which elements belong in $%
\llbracket 1;\mathsf{N} \rrbracket $ \\
$\mathbb{S}$ & $\rightarrow$ the set of all strategies \\
$\mathbf{C}_n^k$ & $\rightarrow$ the binomial coefficient ${n \choose k} = \frac{n!}{k!(n-k)!}$\\
$\oplus$ & $\rightarrow$ bitwise exclusive or \\
$\ll \text{and} \gg$ & $\rightarrow$ the usual shift operators \\
$(\mathcal{X}, \text{d})$ & $\rightarrow$ a metric space  \\
$LCM(a, b)$ & $\rightarrow$ the least common multiple of $a$ and $b$
\end{tabular}

%
%
%
%
\subsection{Chaotic iterations}

\begin{definition}
The set $\mathds{B}$ denoting $\{0,1\}$, let $f:\mathds{B}^{\mathsf{N}%
}\longrightarrow \mathds{B}^{\mathsf{N}}$ be an ``iteration'' function and $S\in \mathbb{S}
$ be a chaotic strategy. Then, the so-called \emph{chaotic iterations} are defined by $x^0\in \mathds{B}^{\mathsf{N}}$, and
\begin{equation}
\begin{array}{l}
\forall n\in \mathds{N}^{\ast },\forall i\in \llbracket1;\mathsf{N}\rrbracket%
,x_i^n=\left\{
\begin{array}{l}
x_i^{n-1}~~~~~\text{if}~S^n\neq i \\
f(x^{n-1})_{S^n}~\text{if}~S^n=i.\end{array} \right. \end{array}
\end{equation}
\end{definition}
In other words, at the $n^{th}$ iteration, only the $S^{n}-$th cell is
\textquotedblleft iterated\textquotedblright.

\subsection{The CIPRNG family}

\subsubsection{Old CIPRNG}

Let $\mathsf{N} = 4$. Some chaotic iterations are fulfilled to generate a sequence $\left(x^n\right)_{n\in\mathds{N}} \in \left(\mathds{B}^4\right)^\mathds{N}$ of Boolean vectors: the successive states of the iterated system. Some of these vectors are randomly extracted and their components constitute our pseudorandom bit flow~\cite{bgw09:ip}.
Chaotic iterations are realized as follows. Initial state $x^0 \in \mathds{B}^4$ is a Boolean vector taken as a seed and chaotic strategy $\left(S^n\right)_{n\in\mathds{N}}\in \llbracket 1, 4 \rrbracket^\mathds{N}$ is constructed with $PRNG_2$. Lastly, iterate function $f$ is the vectorial Boolean negation.
At each iteration, only the $S^n$-th component of state $x^n$ is updated. Finally, some $x^n$ are selected by a sequence $m^n$, provided by a second generator $PRNG_1$, as the pseudorandom bit sequence of our generator.

The basic design procedure of the Old CI generator is summed up in Algorithm~\ref{Chaotic iteration}.
The internal state is $x$, the output array is $r$. $a$ and $b$ are those computed by $PRNG_1$ and $PRNG_2$.

\begin{algorithm}
\textbf{Input:} the internal state $x$ (an array of 4-bit words)\\
\textbf{Output:} an array $r$ of 4-bit words
\begin{algorithmic}[1]

\STATE$a\leftarrow{PRNG_1()}$;
\STATE$m\leftarrow{a~mod~2+13}$;
\WHILE{$i=0,\dots,m$}
\STATE$b\leftarrow{PRNG_2()}$;
\STATE$S\leftarrow{b~mod~4}$;
\STATE$x_S\leftarrow{ \overline{x_S}}$;
\ENDWHILE
\STATE$r\leftarrow{x}$;
\STATE return $r$;
\medskip
\caption{An arbitrary round of the old CI generator}
\label{Chaotic iteration}
\end{algorithmic}
\end{algorithm}

\subsubsection{New CIPRNG}

The New CI generator is designed by the following process~\cite{bg10:ip}. First of all, some chaotic iterations have to be done to generate a sequence $\left(x^n\right)_{n\in\mathds{N}} \in \left(\mathds{B}^{32}\right)^\mathds{N}$ of Boolean vectors, which are the successive states of the iterated system. Some of these vectors will be randomly extracted and our pseudo-random bit flow will be constituted by their components. Such chaotic iterations are realized as follows. Initial state $x^0 \in \mathds{B}^{32}$ is a Boolean vector taken as a seed and chaotic strategy $\left(S^n\right)_{n\in\mathds{N}}\in \llbracket 1, 32 \rrbracket^\mathds{N}$ is
an \emph{irregular decimation} of $PRNG_2$ sequence, as described in Algorithm~\ref{Chaotic iteration1}.

Another time, at each iteration, only the $S^n$-th component of state $x^n$ is updated, as follows: $x_i^n = x_i^{n-1}$ if $i \neq S^n$, else $x_i^n = \overline{x_i^{n-1}}$.
Finally, some $x^n$ are selected
by a sequence $m^n$ as the pseudo-random bit sequence of our generator.
$(m^n)_{n \in \mathds{N}} \in \mathcal{M}^\mathds{N}$ is computed from $PRNG_1$, where $\mathcal{M}\subset \mathds{N}^*$ is a finite nonempty set of integers.

The basic design procedure of the New CI generator is summarized in Algorithm~\ref{Chaotic iteration1}.
The internal state is $x$, the output state is $r$. $a$ and $b$ are those computed by the two input
PRNGs. Lastly, the value $g_1(a)$ is an integer defined as in Eq.~\ref{Formula}.

\begin{equation}
\label{Formula}
m^n = g_1(y^n)=
\left\{
\begin{array}{l}
0 \text{ if }0 \leqslant{y^n}<{C^0_{32}},\\
1 \text{ if }{C^0_{32}} \leqslant{y^n}<\sum_{i=0}^1{C^i_{32}},\\
2 \text{ if }\sum_{i=0}^1{C^i_{32}} \leqslant{y^n}<\sum_{i=0}^2{C^i_{32}},\\
\vdots~~~~~ ~~\vdots~~~ ~~~~\\
N \text{ if }\sum_{i=0}^{N-1}{C^i_{32}}\leqslant{y^n}<1.\\
\end{array}
\right.
\end{equation}

\begin{algorithm}
\textbf{Input:} the internal state $x$ (32 bits)\\
\textbf{Output:} a state $r$ of 32 bits
\begin{algorithmic}[1]
\FOR{$i=0,\dots,N$}
{
\STATE$d_i\leftarrow{0}$\;
}
\ENDFOR
\STATE$a\leftarrow{PRNG_1()}$\;
\STATE$m\leftarrow{f(a)}$\;
\STATE$k\leftarrow{m}$\;
\WHILE{$i=0,\dots,k$}

\STATE$b\leftarrow{PRNG_2()~mod~\mathsf{N}}$\;
\STATE$S\leftarrow{b}$\;
    \IF{$d_S=0$}
    {
\STATE      $x_S\leftarrow{ \overline{x_S}}$\;
\STATE      $d_S\leftarrow{1}$\;

    }
    \ELSIF{$d_S=1$}
    {
\STATE      $k\leftarrow{ k+1}$\;
    }\ENDIF
\ENDWHILE\\
\STATE $r\leftarrow{x}$\;
\STATE return $r$\;
\medskip
\caption{An arbitrary round of the new CI generator}
\label{Chaotic iteration1}
\end{algorithmic}
\end{algorithm}

\subsubsection{Xor CIPRNG}

Instead of updating only one cell at each iteration as Old CI and New CI, we can try to choose a
subset of components and to update them together. Such an attempt leads
to a kind of merger of the two random sequences. When the updating function is the vectorial negation,
this algorithm can be rewritten as follows~\cite{arxivRCCGPCH}:

\begin{equation}
\left\{
\begin{array}{l}
x^0 \in \llbracket 0, 2^\mathsf{N}-1 \rrbracket, S \in \llbracket 0, 2^\mathsf{N}-1 \rrbracket^\mathds{N} \\
\forall n \in \mathds{N}^*, x^n = x^{n-1} \oplus S^n,
\end{array}
\right.
\label{equation Oplus}
\end{equation}

The single basic component presented in Eq.~\ref{equation Oplus} is of
ordinary use as a good elementary brick in various PRNGs. It corresponds
to the discrete dynamical system in chaotic iterations.

\section{About some Well-known PRNGs}
\label{The generation of pseudo-random sequence}

\subsection{Introduction}

Knowing that there is no universal generator, it is strongly recommended to test a stochastic application with a large set of different PRNGs~\cite{DavidRC2003643}. They can be classified in four major classes: linear generators, lagged generators, inversive generators, and mix generators:
\begin{itemize}
 \item \textbf{Linear generators}, defined by a linear recurrence, are the most commonly analyzed and utilized generators. The main linear generators are LCGs and MLCG.
 \item \textbf{Lagged generators} have a general recursive formula that use various previously computed terms in the determination of the new sequence value.
 \item \textbf{Inversive congruential generators} form a recent class of generators that are based on the principle of congruential inversion.
 \item \textbf{Mixed generators} result from the need for sequences of better and better quality, or at least longer periods. This has led to mix different types of PRNGs, as follows: $x^i=y^i\oplus z^i$
\end{itemize}

\begin{figure}
\centering
\includegraphics[width=3.5in]{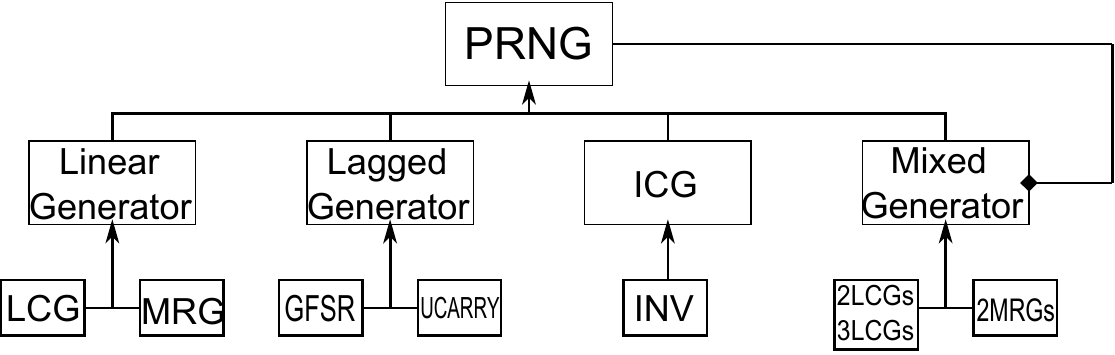}
\DeclareGraphicsExtensions.
\caption{Ontological class hierarchy of PRNGs}
\label{Ontological class hierarchy of RNGs}
\end{figure}

For instance, inversive generators are very interesting for verifying simulation results obtained with a linear congruential generator (LCG),
because their internal structure and correlation behavior strongly differs from what LCGs produce.
Since these generators have revealed several issues, some scientists refrain from using them.
In what follows, chaotic properties will be added to these PRNGs, leading to noticeable improvements observed by statistical test.
Let us firstly explain with more details the generators studied in this research work (for a synthetic view, see Fig.~\ref{Ontological class hierarchy of RNGs}).

\subsection{Details of some Existing Generators}

Here are the modules of PRNGs we have chosen to experiment.

\subsubsection{LCG}
This PRNG implements either the simple or the combined linear congruency generator (LCGs). The simple LCG is defined by the recurrence:
\begin{equation}
x^n = (ax^{n-1} + c)~mod~m
\label{LCG}
\end{equation}
where $a$, $c$, and $x^0$ must be, among other things, non-negative and less than $m$~\cite{testU01}. In what follows, 2LCGs and 3LCGs refer as two (resp. three) combinations of such LCGs.
For further details, see~\cite{combined_lcg}.

\subsubsection{MRG}
This module implements multiple recursive generators (MRGs), based on a linear recurrence of order $k$, modulo $m$~\cite{testU01}:
\begin{equation}
x^n = (a^1x^{n-1}+~...~+a^kx^{n-k})~mod~m
\label{MRG}
\end{equation}
Combination of two MRGs (referred as 2MRGs) is also be used in this paper.

\subsubsection{UCARRY}
Generators based on linear recurrences with carry are implemented in this module. This includes the add-with-carry (AWC) generator, based on the recurrence:
\begin{equation}
\label{AWC}
\begin{array}{l}
x^n = (x^{n-r} + x^{n-s} + c^{n-1})~mod~m, \\
c^n= (x^{n-r} + x^{n-s} + c^{n-1}) / m, \end{array}\end{equation}
the SWB generator, having the recurrence:
\begin{equation}
\label{SWB}
\begin{array}{l}
x^n = (x^{n-r} - x^{n-s} - c^{n-1})~mod~m, \\
c^n=\left\{
\begin{array}{l}
1 ~~~~~\text{if}~ (x^{i-r} - x^{i-s} - c^{i-1})<0\\
0 ~~~~~\text{else},\end{array} \right. \end{array}\end{equation}
and the SWC generator designed by R. Couture, which is based on the following recurrence:
\begin{equation}
\label{SWC}
\begin{array}{l}
x^n = (a^1x^{n-1} \oplus ~...~ \oplus a^rx^{n-r} \oplus c^{n-1}) ~ mod ~ 2^w, \\
c^n = (a^1x^{n-1} \oplus ~...~ \oplus a^rx^{n-r} \oplus c^{n-1}) ~ / ~ 2^w. \end{array}\end{equation}

\subsubsection{GFSR}
This module implements the generalized feedback shift register (GFSR) generator, that is:
\begin{equation}
x^n = x^{n-r} \oplus x^{n-k}
\label{GFSR}
\end{equation}


\subsubsection{INV}
Finally, this module implements the nonlinear inversive generator, as defined in~\cite{testU01}, which is:

\begin{equation}
\label{INV}
\begin{array}{l}
x^n=\left\{
\begin{array}{ll}
(a^1 + a^2 / z^{n-1})~mod~m & \text{if}~ z^{n-1} \neq 0 \\
a^1 & \text{if}~  z^{n-1} = 0 .\end{array} \right. \end{array}\end{equation}

\section{statistical tests}
\label{Security analysis}

Considering the properties of binary random sequences, various statistical tests can be designed to evaluate the assertion that the sequence is generated by a perfectly random source. We have performed some statistical tests for the CIPRNGs proposed here. These tests include NIST suite~\cite{ANDREW2008} and DieHARD battery of tests~\cite{DieHARD}. For completeness and for reference, we give in the following subsection a brief description of each of the aforementioned tests.

\subsection{NIST statistical tests suite}

Among the numerous standard tests for pseudo-randomness, a convincing way to show the randomness of the produced sequences is to confront them to the NIST (National Institute of  Standards and Technology) statistical tests, being an up-to-date tests suite proposed by the Information Technology Laboratory (ITL). A new version of the Statistical tests suite has been released in August 11, 2010.

The NIST tests suite SP 800-22 is a statistical package consisting of 15 tests. They were developed to test the randomness of binary sequences produced by hardware or software based cryptographic pseudorandom number generators. These tests focus on a variety of different types of non-randomness that could exist in a sequence.

For each statistical test, a set of $P-values$ (corresponding to the set of sequences) is produced.
The interpretation of empirical results can be conducted in various ways.
In this paper, the examination of the distribution of P-values to check for uniformity ($ P-value_{T}$) is used.
The distribution of $P-values$ is examined to ensure uniformity.
If $P-value_{T} \geqslant 0.0001$, then the sequences can be considered to be uniformly distributed.

In our experiments, 100 sequences (s = 100), each with 1,000,000-bit long, are generated and tested. If the $P-value_{T}$ of any test is smaller than 0.0001, the sequences are considered to be not good enough and the generating algorithm is not suitable for usage.

\subsection{DieHARD battery of tests}
The DieHARD battery of tests has been the most sophisticated standard for over a decade. Because of the stringent requirements in the DieHARD tests suite, a generator passing this battery of
tests can be considered good as a rule of thumb.

The DieHARD battery of tests consists of 18 different independent statistical tests. This collection
 of tests is based on assessing the randomness of bits comprising 32-bit integers obtained from
a random number generator. Each test requires $2^{23}$ 32-bit integers in order to run the full set
of tests. Most of the tests in DieHARD return a $P-value$, which should be uniform on $[0,1)$ if the input file
contains truly independent random bits.  These $P-values$ are obtained by
$P=F(X)$, where $F$ is the assumed distribution of the sample random variable $X$ (often normal).
But that assumed $F$ is just an asymptotic approximation, for which the fit will be worst
in the tails. Thus occasional $P-values$ near 0 or 1, such as 0.0012 or 0.9983, can occur.
An individual test is considered to be failed if the $P-value$ approaches 1 closely, for example $P>0.9999$.

\section{Results and discussion}
\label{Results and discussion}
\begin{table*}
\renewcommand{\arraystretch}{1.3}
\caption{NIST and DieHARD tests suite passing rates for PRNGs without CI}
\label{NIST and DieHARD tests suite passing rate the for PRNGs without CI}
\centering
  \begin{tabular}{|l||c|c|c|c|c|c|c|c|c|c|}
    \hline\hline
Types of PRNGs & \multicolumn{2}{c|}{Linear PRNGs} & \multicolumn{4}{c|}{Lagged PRNGs} & \multicolumn{1}{c|}{ICG PRNGs} & \multicolumn{3}{c|}{Mixed PRNGs}\\ \hline
\backslashbox{\textbf{$Tests$}} {\textbf{$PRNG$}} & LCG& MRG& AWC & SWB  & SWC & GFSR & INV & LCG2& LCG3& MRG2 \\ \hline
NIST & 11/15 & 14/15 &\textbf{15/15} & \textbf{15/15}   & 14/15 & 14/15  & 14/15 & 14/15& 14/15& 14/15 \\ \hline
DieHARD & 16/18 & 16/18 & 15/18 & 16/18 & \textbf{18/18} & 16/18 & 16/18 & 16/18& 16/18& 16/18\\ \hline
\end{tabular}
\end{table*}

Table~\ref{NIST and DieHARD tests suite passing rate the for PRNGs without CI} shows the results on the batteries recalled above, indicating that almost all the PRNGs cannot pass all their tests. In other words, the statistical quality of these PRNGs cannot fulfill the up-to-date standards presented previously. We will show that the CIPRNG can solve this issue.

To illustrate the effects of this CIPRNG in detail, experiments will be divided in three parts:
\begin{enumerate}
  \item \textbf{Single CIPRNG}: The PRNGs involved in CI computing are of the same category.
  \item \textbf{Mixed CIPRNG}: Two different types of PRNGs are mixed during the chaotic iterations process.
  \item \textbf{Multiple CIPRNG}: The generator is obtained by repeating the composition of the iteration function as follows: $x^0\in \mathds{B}^{\mathsf{N}}$, and $\forall n\in \mathds{N}^{\ast },\forall i\in \llbracket1;\mathsf{N}\rrbracket,$
\begin{equation}
\begin{array}{l}
x_i^n=\left\{
\begin{array}{l}
x_i^{n-1}~~~~~\text{if}~S^n\neq i \\
\forall j\in \llbracket1;\mathsf{m}\rrbracket,f^m(x^{n-1})_{S^{nm+j}}~\text{if}~S^{nm+j}=i.\end{array} \right. \end{array}
\end{equation}
$m$ is called the \emph{functional power}.
\end{enumerate}

We have performed statistical analysis of each of the aforementioned CIPRNGs.
The results are reproduced in Tables~\ref{NIST and DieHARD tests suite passing rate the for PRNGs without CI} and \ref{NIST and DieHARD tests suite passing rate the for single CIPRNGs}.
The scores written in boldface indicate that all the tests have been passed successfully, whereas an asterisk ``*'' means that the considered passing rate has been improved.

\subsection{Tests based on the Single CIPRNG}

\begin{table*}
\renewcommand{\arraystretch}{1.3}
\caption{NIST and DieHARD tests suite passing rates for PRNGs with CI}
\label{NIST and DieHARD tests suite passing rate the for single CIPRNGs}
\centering
  \begin{tabular}{|l||c|c|c|c|c|c|c|c|c|c|c|c|}
    \hline
Types of PRNGs & \multicolumn{2}{c|}{Linear PRNGs} & \multicolumn{4}{c|}{Lagged PRNGs} & \multicolumn{1}{c|}{ICG PRNGs} & \multicolumn{3}{c|}{Mixed PRNGs}\\ \hline
\backslashbox{\textbf{$Tests$}} {\textbf{$Single~CIPRNG$}} & LCG  & MRG & AWC & SWB & SWC & GFSR & INV& LCG2 & LCG3& MRG2 \\ \hline\hline
Old CIPRNG\\ \hline \hline
NIST & \textbf{15/15} *  & \textbf{15/15} * & \textbf{15/15}   & \textbf{15/15}   & \textbf{15/15} * & \textbf{15/15} * & \textbf{15/15} *& \textbf{15/15} * & \textbf{15/15} * & \textbf{15/15} \\ \hline
DieHARD & \textbf{18/18} *  & \textbf{18/18} * & \textbf{18/18} *  & \textbf{18/18} *  & \textbf{18/18}  & \textbf{18/18} * & \textbf{18/18} *& \textbf{18/18} * & \textbf{18/18} *& \textbf{18/18} * \\ \hline
New CIPRNG\\ \hline \hline
NIST & \textbf{15/15} *  & \textbf{15/15} * & \textbf{15/15}   & \textbf{15/15}  & \textbf{15/15} * & \textbf{15/15} * & \textbf{15/15} *& \textbf{15/15} * & \textbf{15/15} * & \textbf{15/15} \\ \hline
DieHARD & \textbf{18/18} *  & \textbf{18/18} * & \textbf{18/18} * & \textbf{18/18} * & \textbf{18/18}  & \textbf{18/18} * & \textbf{18/18} * & \textbf{18/18} * & \textbf{18/18} *& \textbf{18/18} *\\ \hline
Xor CIPRNG\\ \hline\hline
NIST & 14/15*& \textbf{15/15} *   & \textbf{15/15}   & \textbf{15/15}   & 14/15 & \textbf{15/15} * & 14/15& \textbf{15/15} * & \textbf{15/15} *& \textbf{15/15}  \\ \hline
DieHARD & 16/18 & 16/18 & 17/18* & \textbf{18/18} * & \textbf{18/18}  & \textbf{18/18} * & 16/18 & 16/18 & 16/18& 16/18\\ \hline
\end{tabular}
\end{table*}

The statistical tests results of the PRNGs using the single CIPRNG method are given in Table~\ref{NIST and DieHARD tests suite passing rate the for single CIPRNGs}.
We can observe that, except for the Xor CIPRNG, all of the CIPRNGs have passed the 15 tests of the NIST battery and the 18 tests of the DieHARD one.
Moreover, considering these scores, we can deduce that both the single Old CIPRNG and the single New CIPRNG are relatively steadier than the single Xor CIPRNG approach, when applying them to different PRNGs.
However, the Xor CIPRNG is obviously the fastest approach to generate a CI random sequence, and it still improves the statistical properties relative to each generator taken alone, although the test values are not as good as desired.

Therefore, all of these three ways are interesting, for different reasons, in the production of pseudorandom numbers and,
on the whole, the single CIPRNG method can be considered to adapt to or improve all kinds of PRNGs.

To have a realization of the Xor CIPRNG that can pass all the tests embedded into the NIST battery, the Xor CIPRNG with multiple functional powers are investigated in Section~\ref{Tests based on Multiple CIPRNG}.

\subsection{Tests based on the Mixed CIPRNG}

To compare the previous approach with the CIPRNG design that uses a Mixed CIPRNG, we have taken into account the same inputted generators than in the previous section.
These inputted couples $(PRNG_1,PRNG_2)$ of PRNGs are used in the Mixed approach as follows:
\begin{equation}
\left\{
\begin{array}{l}
x^0 \in \llbracket 0, 2^\mathsf{N}-1 \rrbracket, S \in \llbracket 0, 2^\mathsf{N}-1 \rrbracket^\mathds{N} \\
\forall n \in \mathds{N}^*, x^n = x^{n-1} \oplus PRNG_1\oplus PRNG_2,
\end{array}
\right.
\label{equation Oplus}
\end{equation}

With this Mixed CIPRNG approach, both the Old CIPRNG and New CIPRNG continue to pass all the NIST and DieHARD suites.
In addition, we can see that the PRNGs using a Xor CIPRNG approach can pass more tests than previously.
The main reason of this success is that the Mixed Xor CIPRNG has a longer period.
Indeed, let $n_{P}$ be the period of a PRNG $P$, then the period deduced from the single Xor CIPRNG approach is obviously equal to:
\begin{equation}
n_{SXORCI}=
\left\{
\begin{array}{ll}
n_{P}&\text{if~}x^0=x^{n_{P}}\\
2n_{P}&\text{if~}x^0\neq x^{n_{P}}.\\
\end{array}
\right.
\label{equation Oplus}
\end{equation}

Let us now denote by $n_{P1}$ and $n_{P2}$ the periods of respectively the $PRNG_1$ and $PRNG_2$ generators, then the period of the Mixed Xor CIPRNG will be:
\begin{equation}
n_{XXORCI}=
\left\{
\begin{array}{ll}
LCM(n_{P1},n_{P2})&\text{if~}x^0=x^{LCM(n_{P1},n_{P2})}\\
2LCM(n_{P1},n_{P2})&\text{if~}x^0\neq x^{LCM(n_{P1},n_{P2})}.\\
\end{array}
\right.
\label{equation Oplus}
\end{equation}

In Table~\ref{DieHARD fail mixex CIPRNG}, we only show the results for the Mixed CIPRNGs that cannot pass all DieHARD suites (the NIST tests are all passed). It demonstrates that Mixed Xor CIPRNG involving LCG, MRG, LCG2, LCG3, MRG2, or INV cannot pass the two following tests, namely the ``Matrix Rank 32x32'' and the ``COUNT-THE-1's'' tests contained into the DieHARD battery. Let us recall their definitions:

\begin{itemize}
 \item \textbf{Matrix Rank 32x32.} A random 32x32 binary matrix is formed, each row having a 32-bit random vector. Its rank is an integer that ranges from 0 to 32. Ranks less than 29 must be rare, and their occurences must be pooled with those of rank 29. To achieve the test, ranks of 40,000 such random matrices are obtained, and a chisquare test is performed on counts for ranks 32,31,30 and for ranks $\leq29$.

 \item \textbf{COUNT-THE-1's TEST} Consider the file under test as a stream of bytes (four per  2 bit integer).  Each byte can contain from 0 to 8 1's, with probabilities 1,8,28,56,70,56,28,8,1 over 256.  Now let the stream of bytes provide a string of overlapping  5-letter words, each ``letter'' taking values A,B,C,D,E. The letters are determined by the number of 1's in a byte: 0,1, or 2 yield A, 3 yields B, 4 yields C, 5 yields D and 6,7, or 8 yield E. Thus we have a monkey at a typewriter hitting five keys with various probabilities (37,56,70,56,37 over 256).  There are $5^5$ possible 5-letter words, and from a string of 256,000 (over-lapping) 5-letter words, counts are made on the frequencies for each word.   The quadratic form in the weak inverse of the covariance matrix of the cell counts provides a chisquare test: Q5-Q4, the difference of the naive Pearson sums of $(OBS-EXP)^2/EXP$ on counts for 5- and 4-letter cell counts.
\end{itemize}

The reason of these fails is that the output of LCG, LCG2, LCG3, MRG, and MRG2 under the experiments are in 31-bit. Compare with the Single CIPRNG, using different PRNGs to build CIPRNG seems more efficient in improving random number quality (mixed Xor CI can 100\% pass NIST, but single cannot).

\begin{table*}
\renewcommand{\arraystretch}{1.3}
\caption{Scores of mixed Xor CIPRNGs when considering the DieHARD battery}
\label{DieHARD fail mixex CIPRNG}
\centering
  \begin{tabular}{|l||c|c|c|c|c|c|}
    \hline
\backslashbox{\textbf{$PRNG_1$}} {\textbf{$PRNG_0$}} & LCG & MRG & INV & LCG2 & LCG3 & MRG2 \\ \hline\hline
LCG  &\backslashbox{} {} &16/18&16/18 &16/18 &16/18 &16/18\\ \hline
MRG &16/18 &\backslashbox{} {} &16/18&16/18 &16/18  &16/18\\ \hline
INV &16/18 &16/18&\backslashbox{} {} &16/18 &16/18&16/18    \\ \hline
LCG2  &16/18 &16/18 &16/18 &\backslashbox{} {}  &16/18&16/18\\ \hline
LCG3  &16/18 &16/18 &16/18&16/18&\backslashbox{} {} &16/18\\ \hline
MRG2 &16/18  &16/18 &16/18&16/18 &16/18 &\backslashbox{} {}  \\ \hline
\end{tabular}
\end{table*}

\subsection{Tests based on the Multiple CIPRNG}
\label{Tests based on Multiple CIPRNG}

Until now, the combination of at most two input PRNGs has been investigated.
We now regard the possibility to use a larger number of generators to improve the statistics of the generated pseudorandom numbers, leading to the multiple functional power approach.
For the CIPRNGs which have already pass both the NIST and DieHARD suites with 2 inputted PRNGs (all the Old and New CIPRNGs, and some of the Xor CIPRNGs), it is not meaningful to consider their adaption of this multiple CIPRNG method, hence only the Multiple Xor CIPRNGs, having the following form, will be investigated.
\begin{equation}
\left\{
\begin{array}{l}
x^0 \in \llbracket 0, 2^\mathsf{N}-1 \rrbracket, S \in \llbracket 0, 2^\mathsf{N}-1 \rrbracket^\mathds{N} \\
\forall n \in \mathds{N}^*, x^n = x^{n-1} \oplus S^{nm}\oplus S^{nm+1}\ldots \oplus S^{nm+m-1} ,
\end{array}
\right.
\label{equation Oplus}
\end{equation}

The question is now to determine the value of the threshold $m$ (the functional power) making the multiple CIPRNG being able to pass the whole NIST battery.
Such a question is answered in Table~\ref{threshold}.

\begin{table*}
\renewcommand{\arraystretch}{1.3}
\caption{Functional power $m$ making it possible to pass the whole NIST battery}
\label{threshold}
\centering
  \begin{tabular}{|l||c|c|c|c|c|c|c|c|}
    \hline
Inputted $PRNG$ & LCG & MRG & SWC & GFSR & INV& LCG2 & LCG3  & MRG2 \\ \hline\hline
Threshold  value $m$& 19 & 7  & 2& 1 & 11& 9& 3& 4\\ \hline\hline
\end{tabular}
\end{table*}

\subsection{Results Summary}

We can summarize the obtained results as follows.
\begin{enumerate}
\item The CIPRNG method is able to improve the statistical properties of a large variety of PRNGs.
\item Using different PRNGs in the CIPRNG approach is better than considering several instances of one unique PRNG.
\item The statistical quality of the outputs increases with the functional power $m$.
\end{enumerate}

\section{Conclusion and Future Work}

In this paper, we first have formalized the CI methods that has been already presented in  previous Internet conferences.
These CI methods are based on iterations that have been topologically proven as chaotic.
Then 10 usual PRNGs covering all kinds of generators have been applied, and the NIST and DieHARD batteries have been tested.
Analyses show that PRNGs using the CIPRNG methods do not only inherit the chaotic properties of the
CI iterations, they also have improvements of their statistics.
This is why CIPRNG techniques should be considered as post-treatments on pseudorandom number generators to improve both their randomness and security.

In future work, we will try to enlarge this study, by considering a larger variety of tests.
The CIPRNG's chaotic behavior will be deepened by using some specific tools provided by the mathematical theory of chaos.
Finally, a large variety of Internet usages, as cryptography and data hiding, will be considered for applications.



%


\bibliographystyle{plain}
\bibliography{mabase}
\end{document}